\documentclass[preprintnumbers,amsmath,amssymb,twocolumn]{revtex4}
\usepackage[dvips,colorlinks=true,citecolor=red,filecolor=green,linkcolor=blue,pdfnewwindow=true]{hyperref}
\usepackage{graphicx}
\usepackage[title]{appendix}
\usepackage{color}

\begin{document}

\title{Evolution of language driven by social dynamics}
\author{Moirangthem Shubhakanta Singh$^1$ and R.K. Brojen Singh$^2$}
\email{brojen@jnu.ac.in (Corresponding author)}
\affiliation{$^1$ Department of Physics, Manipur University, Canchipur-795003, Manipur, India.\\
$^2$School of Computational \& Integrative Sciences, Jawaharlal Nehru University, New Delhi-110067, India.}
\affiliation{}

\begin{abstract}
{\noindent}The survival of endangered languages in complex language competition depends on socio-cultural status and honour endowed (by itself and by the other) among them. The restriction in the endorsement of this honour leads to language extinction of one language, and rise of the other. Endorsing proper mutual honour each other trigger the co-existence of language speakers and can save both languages from extinction. The lost of respect to each other drives the death of both languages. We found a minimal or critical mutual honour (a=0.9635) which protects the two languages from extinction. The increase in mutual honour from this minimal value allows increase in the populations of the two languages speakers. The state of co-existence of competing languages abolishes the concept of minority and majority in language competition which can be obtained by mutual honour. Further, excess biased honour to a particular language (minority or majority) force the language to extinct. In mean-field approximation of language competition, magnetization parameter can be taken as an indicator of survival of a language.
\end{abstract}


\maketitle

\section{Introduction}
{\noindent}Language carries socio-cultural status of the speakers \cite{fish}, and therefore trigger competition among the languages \cite{mine} which leads to the extinction of some languages and dominance by some other languages \cite{abra}. This language competition could be affected by various factors, namely, decisions in social consensus \cite{migu}, interaction of heterogeneous agents in the society \cite{wang} and many other factors. The two language competition modeled by Abrams and Strogatz \cite{abra} proposed the diversity of languages caused by the competition, which endangers one of the languages leading to extinction. The model has limitations in describing dynamics of individual speakers considering only the behavior of population as a whole. The model was extended to bilanguage competition, which takes into account some individual speakers of both languages in the dynamics \cite{mine}. The incorporation of bilangual speakers as a community in the two-language competition model was then proposed which could lead to the multi-language modeling \cite{mira}. On the other hand there have language models where the two language competition was studied as predator-prey model \cite{pina}.

People carry language, and language spread out from the inhabitant region through interaction, which is proportional to the population size of that language speakers \cite{gong}. Competition between any two languages is greatly influenced by various factors, namely, ability to influence on a language to the others, inheriting capacity of a language to the people, social and cultural status of the language. Bilangualism model is one which incorporates bilangual speakers in the competition of two mutually intelligible or genetically similar languages allowing to maintain the endangered language \cite{mira}. It was proposed, within this model, that even though there is no endangered monolingual speakers, the existence of bilangual speakers preserve the endangered language preventing from extinction \cite{mine}. However, if there is no linguistic advantage of the two competing languages, the disappearance of endangered language speakers will lead to extinction of the language due to conversion of bilangual speakers to the other existing language \cite{mira,mine}. In these models, existence of bilangual speakers in language competition is an indicator of preserving endangered languages from extinction.

Language competition is greatly influence by social status, cultural inheritance and hounor given among the speakers. These influencing factors are so strong that minority languages, which are far different from majority language, can able to survive if these factors are provided among them. We address these issues in this work in order to show the probable preventive measures to be taken up to save endangered languages from extinction. We studied the bilangual model \cite{mine,mira}, which is an extended version of two language competition due to Abrams and Strogatz \cite{abra}, to understand the fundamental roles of the factors in language dynamics and evolution.
\begin{figure}
\label{fig1}
\begin{center}
\includegraphics[height=8.0cm,width=8.0cm]{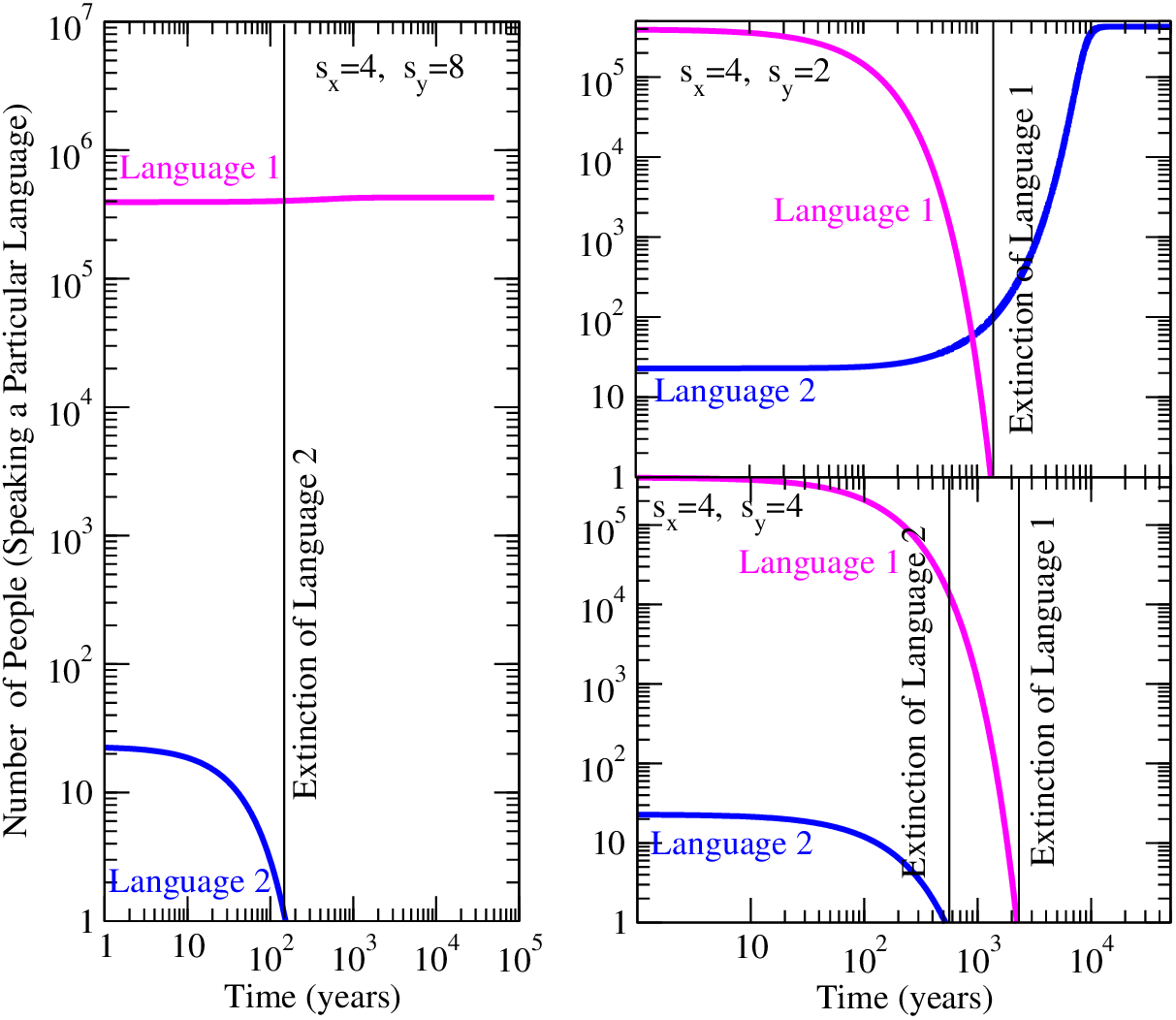}
\caption{The dynamics of two competing languages. Extinction and survival of languages due to biased social status to the languages (left panel) for $c_{XB}=0.035,c_{YB}=0.035,c_{BX}=1.0,c_{BY}=1.0$. Switching of language survival due to change of social status (upper right panel). Low social status drives both languages to extinct (lower right panel).} 
\end{center}
\end{figure}

\section{\bf Bilangual model of two language competition}
The two language competition model due to Abrams and Strogatz \cite{abra} was expended by Minett and Wang \cite{mine} by incorporating the emergence of bilangual speakers from the two language speakers. If $X$ and $Y$ are two competing languages which permits to emerge a group of people $B$ who speak both the languages, the competition allows these basic transitions in due course: $X\rightarrow Y$, $Y\rightarrow X$, $X\rightarrow B$, $B\rightarrow X$, $Y\rightarrow B$ and $B\rightarrow Y$ \cite{mine}. The first two transition can be taken as rare transitions \cite{mine}, and therefore it can be assumed that only the last four transitions are trivially observed transitions during competition \cite{wang}. If $N$ is the population of the monolinguals and bilangual speakers, then $X+Y+B=N$, so that one can define $x=X/N$, $y=Y/N$ and $bB/N$ such that $x+y+b=1$. The transition probability of a transition from state $B$ to $X$ is given by the power law, $P_{BX}=\mu c_{BX}s_Xx^u$, where, $c_{BX}$ is attractiveness of the bilangual speakers towards $X$ speakers, $s_X$ is status of the language $X$ and $\mu$ is the mortality rate of $B$, i.e. replacing rate of $B$ by $X$. Similarly, the probability that the transition $X\rightarrow B$ occurs is given by, $P_{XB}=(1-\mu)c_{XB}s_Yy^u$, where, $s_X+s_Y=1$. Then the time evolutions of $x$ and $y$ are given by the net change in $x$ and $y$ at state $X$ and $Y$ respectively given by,
\begin{eqnarray}
\frac{dx}{dt}&=&zP_{BX}-xP_{XB}\nonumber\\
&=&\mu c_{BX}s_X(1-x-y)x^u-(1-\mu)c_{XB}s_Yxy^u\\
\frac{dy}{dt}&=&zP_{BY}-yP_{YB}\nonumber\\
&=&\mu c_{BY}s_Y(1-x-y)y^u-(1-\mu)c_{YB}s_Xyx^u
\end{eqnarray}
The critical points of the above coupled differential equations (1) and (2) from the conditions, $\frac{dx}{dt}=0,\frac{dy}{dt}=0$. The critical points ($x^*,y^*$) are calculated from this conditions, and found to be (0,0) as one critical point, and other can be obtained from the equations and the following expression,
\begin{eqnarray}
x^*=\Lambda y^*;~~\Lambda=\left(\frac{c_{XB}c_{BY}s_Y^2}{c_{BX}c_{YB}s_X^2}\right)^{\frac{1}{2u-1}}
\end{eqnarray}
This expression and the two equations from $\frac{dx}{dt}=0,\frac{dy}{dt}=0$ lead to the expressions for $x^*$ and $y^*$ given by,
\begin{eqnarray}
x^*&=&\frac{\Lambda}{1+\Lambda+\Gamma}\\
y^*&=&\frac{1}{1+\Lambda+\Gamma}\\
&&\Gamma=\Lambda\left(\frac{1}{\mu}-1\right)\frac{c_{XB}s_Y}{\mu c_{BX}s_X}\nonumber
\end{eqnarray}
The critical point given by (4) and (5) depends on language bias, social status and hounor provided. 
\begin{figure}
\label{fig2}
\begin{center}
\includegraphics[height=8.0cm,width=8.0cm]{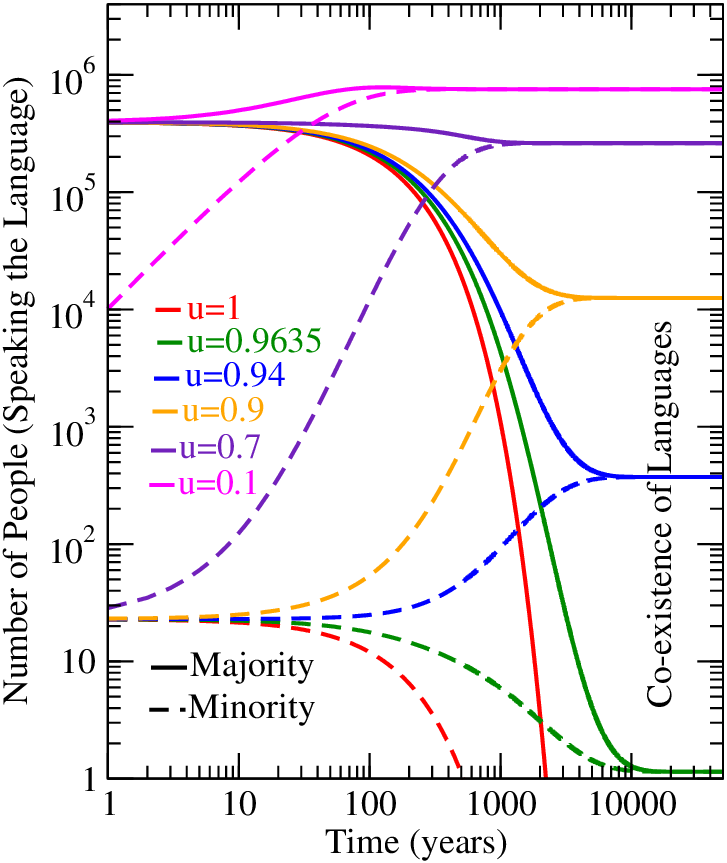}
\caption{Mutual honour endorsed to two competing languages allows to co-existence of both the languages. The dynamics co-existence of both the languages as a function of $u$ is shown.} 
\end{center}
\end{figure}
\begin{figure}
\label{fig3}
\begin{center}
\includegraphics[height=8.0cm,width=8.0cm]{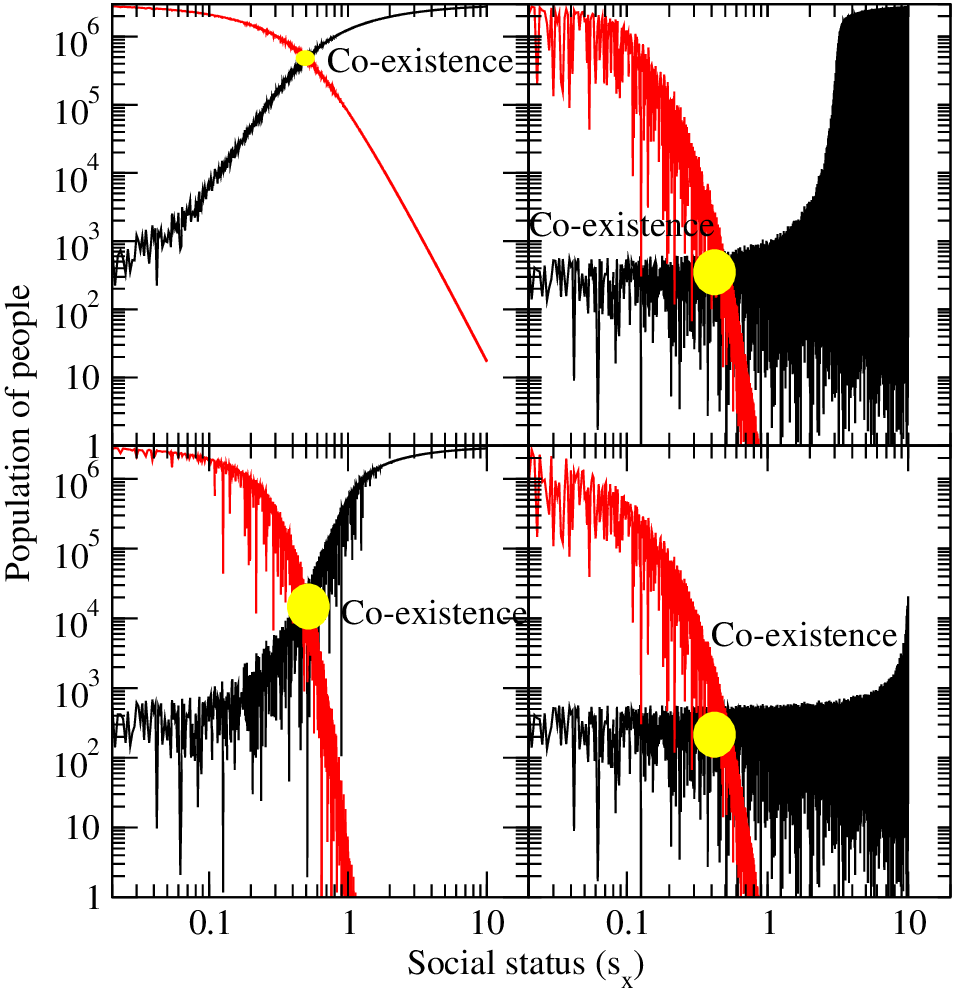}
\caption{Plots of two competing language dynamics as a function of social status of one language $X$ for a fixed value of $s_X$.} 
\end{center}
\end{figure}

\section{Mean-field approximation of language competition}
The infinite population limit allows the society fully connected due to interaction of various language speakers \cite{cast}. In this limit, one can define $m=x-y$, which is closely similar to magnetization parameter. If $\nu$ is a bias parameter, defined by $\nu=s_X-s_Y$, then putting $s_X=s$, we get, $\nu=2s-1$. Then, the time evolution of magnetization parameter, $m$ can be obtained from equations (1) and (2). Similarly, from $x+y+b=1$ we have, $\frac{db}{dt}=-\frac{d}{dt}(x+y)$. The time evolution of $m$ and $b$ are given by,
\begin{eqnarray}
\frac{dm}{dt}&=&\frac{\mu}{2^u}\left[c_{BX}s(1-b+m)^u\right.\nonumber\\
&&\left.-c_{BY}(s-\nu)(1-b-m)^u\right]\nonumber\\
&&+\frac{1-\mu}{2^{u+1}}\left[(1-b)^2-m^2\right]\nonumber\\
&&\times\left[c_{XB}(s-\nu)(1-b-m)^{u-1}\right.\nonumber\\
&&\left.-c_{YB}s(1-b+m)^{u-1}\right]\\
\frac{db}{dt}&=&\frac{\mu}{2^u}\left[c_{BX}s(1-b+m)^u\right.\nonumber\\
&&\left.+c_{BY}(s-\nu)(1-b-m)^u\right]\nonumber\\
&&+\frac{\mu-1}{2^{u+1}}\left[(1-b)^2-m^2\right]\nonumber\\
&&\times\left[c_{XB}(s-\nu)(1-b-m)^{u-1}\right.\nonumber\\
&&\left.+c_{YB}s(1-b+m)^{u-1}\right]
\end{eqnarray}
The critical points $(m^*,b^*)$ of the above differential equations (6) and (7) can be obtained using the conditions $\frac{dm}{dt}=0,~\frac{db}{dt}=0$. The two equations obtained from these two conditions give two critical points, (0,1) and the other critical point is given by,
\begin{eqnarray}
m^*&=&\frac{\mu}{\mu-1}\left[\frac{c_{BY}(s-\nu)}{c_{YB}s}-\frac{c_{BX}s}{c_{XB}(s-\nu)}\right]\\
b^*&=&1-\frac{\mu}{1-\mu}\left[\frac{c_{BX}s}{c_{XB}(s-\nu)}+\frac{c_{BY}(s-\nu)}{c_{YB}s}\right]
\end{eqnarray}
The above critical point is greatly affected by various parameters, such as, status, bias and mortality rate of the languages. The population difference of the two language speakers remains constant which means that there is no competition between the two languages providing harmony by the bilangual speakers.

\section{Results and discussions}
The language dynamics driven by various parameters are presented here by solving the coupled differential equations provided in equations (1), (2), (8) and (8) using standard fourth order Runge-kutta method of numerical integration \cite{pres}. The complicated language dynamics, affected by various social and cultural factors, is studied and provide some of the possible criteria to save endangered languages from extinction.

\subsection{Language dynamics driven by social status}
The social status of the two competing languages, indicated by $s_X$ and $s_Y$, influence greatly the attractiveness of each language leading to the conversion of speakers of one language to the other, and vice versa. If $X$ and $Y$ are minority and majority languages respectively, and the social status endorsed to both the languages has large difference ($s_X=4,~s_Y=8$), then the speakers of minority language will decrease monotonically with time (Fig. 1 left panel) and will be extinct after a particular time. The speakers of majority language will be increased for certain range of time due to conversion of minority to majority language speakers, and then remain stationary (Fig. 1 left panel). The extinction time of the minority language slow if the social status difference, $\Delta s=(s_Y-s_X)\rightarrow 0$ with large $s_X,s_Y$, and minority language comes into existence if $\Delta s=0$.

The social status given in the majority language is smaller as compared to the minority language, $s_X\rangle s_Y$, then the majority language will become extinct and minority language can able to survive (Fig. 1 upper right panel). Further, if $s_X,s_Y\rightarrow small$, then even if $\Delta s=0$, both minority and majority languages will become extinct (Fig. 1 right lower panel). Hence the social status to be given to the competing languages should be given large and same, so that both the language could able to survive.

\subsection{Language driven by mutual honour}
The mutual honour (indicated by parameter $u$) endorsed to each other in language competition plays an important role which can able to save both minority and majority languages from extinction (Fig. 2). Low social status ($s_X=0.4,~s_Y=0.4$) with low mutual honour (large value of $u=1$) given to each other drive both languages to extinct at different times (Fig. 1 and 2). Keeping this social status fixed, decreasing the value of $u$ allows to increase the number of the language speakers. The maximum critical value of $u$ is found to be $u=0.9635$ which can save the both the languages from extinction (non-zero population). The populations of the two languages speakers first decreases upto minimum non-zero populations, and then the two populations become equal and constant as a function of time. After this time of co-existence there is no distinction between majority and minority languages because of the same populations of the two languages. This indicates the co-existence of the two language speakers at this social status, needs minimum honour between the two language speakers which can be used to increase the attractiveness of each individual language, and can able to save the languages from extinction (Fig. 2). 

Now, if the value of $u$ decreases the co-existence of the two languages occurs much earlier, and the numbers of the language speakers are increased. This increase in the populations of two languages at co-existence is due to proper honour to each other.
\begin{figure}
\label{fig4}
\begin{center}
\includegraphics[height=8.0cm,width=8.0cm]{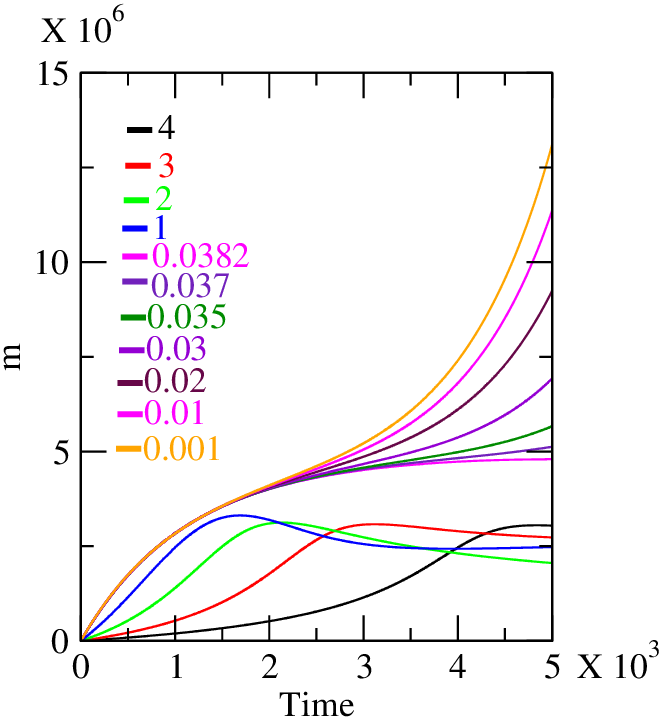}
\caption{The dynamics of magnetization for different values of mutual honour parameter $u$ for parameter values $c_{XB}=0.008,c_{YB}=0.008,c_{BX}=0.035,c_{BY}=0.035, \mu=0.05, \nu=0.3, s=0.35$.} 
\end{center}
\end{figure}

\subsection{Social status can save minority language}
The population of majority and minority language speakers are calculated as a function of $s_X$ for a fixed value of $s_Y$ (Fig. 3). The populations of the speakers of the two languages fluctuate as a function of social status of the minority language ($s_X$), but the population of minority language speakers increases as $s_X$ increases. At the same time the population of the majority language decreases as a function of $s_X$ with large fluctuations due to social fluctuations with time. 

The co-existence of the two languages happens only at few points, and then they are parted away from the co-existence. These co-existing points are the points at which both the languages can be save from extinction. Beyond this/these points one of the languages will be extinct.

\subsection{Magnetization as indicator of language extinction}
The magnetization $m$ is calculated as a function of time for different values of $u$ for fixed values of other parameter values (Fig. 4). Magnetization for large values of $u$ ($u>1$) shows slow variation with time and the behavior become steady state. The dynamics of $m$ for small values of $u$ ($u<0.0382$) become monotonically increased as $u$ decreases. Since $m=x-y$, the monotonic increase in $m$ with time indicates the dominance of one language over the other leading to the extinction of one language. 

The increase in $m$ is due to language competition, and reaching steady state indicates the co-existence of the two languages where conversion of one language speaker to the other and vice versa does not takes place. The decrease in $m$ for $u=2$ as a function of time indicates either both the languages extinct or co-existence of the two languages (same populations of the language speakers).

\section{Conclusion}
Language dynamics is a complicated process, which involves competition, change, evolution and struggle for survival. The dynamics of competing languages is affected by various factors, namely, mutual honour, prestige, social status and carrying culture which can drive the languages at different states, i.e. extinct or survival. In competing two languages, the language which has relatively low social status will quickly extinct and the language having relatively higher social status will able to survive in long run. However, if the social status of both the languages are smaller than a critical value, both the competing languages will become extinct.

The survival of bilangual speakers in two language competition could be a signature of maintaining minority language to survive by them along with majority language. It could be due to the fact that the minority language is preserved by these speakers along with the other language. However, if the status of the minority language is very low as compared to the other, then the bilingual speakers will be attracted and convert quickly towards the higher status language, and the other language will be extinct. The competition between the two languages is done via transition through bilingual speakers. 

The co-existence between two languages is one peculiar state of competing language dynamics which allows to survive both the languages with same population. This is possible only when the mutual honour between each other is increased above a critical value. Hence the survival of both the languages could be possible if the two competing languages endorse mutual honour and respect.

Magnetization in mean-field approximation can be used as an indicator of language survival in language competition. The monotonic increase in magnetization shows the survival of one language and the other language will extinct. The steady state of dynamics of magnetization could be a state of co-existence of the two language. The decrease in the value of this parameter may indicate extinction of both languages or co-existence state. However, the language competition is a complicated process, and there are lots of issues in understanding this process so that one can save endangered languages. The breakdown of a language into a number of dialects, which may become independent languages, and understanding their complicated dynamics need to be investigated to highlight importance of language competition.

\vspace{0.5cm}
\noindent {\bf Acknowledgments} \\
The author thanks the Special Centre for Study of North-East India, Jawaharlal Nehru University for inviting me to present this work and exciting discussion.

\end{document}